\documentclass[12pt,preprint,letterpaper]{aastex}

\received{?? June 2008} \accepted{??} \journalid{337}{1 January 1989}

\begin{document}

\def\K{{\rm\,K}}
\def\cm{{\rm\,cm}}
\def\sec{{\rm\,s}}
\def\ergs{{\rm\,ergs}}

\title{Mass Accretion Rate of Rotating Viscous Accretion Flow}

\author{Myeong-Gu Park}
\affil{Department of Astronomy and Atmospheric Sciences,
                 Kyungpook National University, Daegu, South Korea; and
                 Department of Astrophysical Sciences,
                 Princeton University, Princeton, NJ; mgp@knu.ac.kr}

\shorttitle{Mass Accretion Rate}
\shortauthors{Park}

\begin{abstract}
The mass accretion rate of transonic spherical accretion flow onto
compact objects such as black holes is known as the Bondi accretion
rate, which is determined only by the density and the temperature of
gas at the outer boundary. A rotating accretion flow has angular
momentum, which modifies the flow profile from the spherical Bondi
flow, and hence its mass accretion rate, but most work on disc
accretion has taken the mass flux to be a given with the relation
between that parameter and external conditions left uncertain. Within
the framework of a slim $\alpha$ disk, we have constructed global
solutions of the rotating, viscous hot accretion flow in the
Paczy\'{n}ski-Wiita potential and determined its mass accretion rate as
a function of density, temperature, and angular momentum of gas at the
outer boundary. We find that the low angular momentum flow resembles
the spherical Bondi flow and its mass accretion rate approaches the
Bondi accretion rate for the same density and temperature at the outer
boundary. The high angular momentum flow on the other hand is the
conventional hot accretion disk with advection, but its mass accretion
rate can be significantly smaller than the Bondi accretion rate with
the same boundary conditions. We also find that solutions exist only
within a limited range of dimensionless mass accretion rate $\dot{m}
\equiv \dot{M}/\dot{M}_B$, where $\dot{M}$ is the mass accretion rate
and $\dot{M}_B$ the Bondi accretion rate: When the temperature at the
outer boundary is equal to the virial temperature, solutions exist only
for $0.05 \la \dot{m} \le 1$ when $\alpha=0.01$. We also find that the
dimensionless mass accretion rate is roughly independent of the radius
of the outer boundary but inversely proportional to the angular
momentum at the outer boundary and proportional to the viscosity
parameter, $\dot{m} \simeq 9.0\ \alpha \lambda^{-1}$ when $0.1 \la
\dot{m} \la 1$, where the dimensionless angular momentum measure
$\lambda \equiv l_{out}/l_B$ is the specific angular momentum of gas at
the outer boundary $l_{out}$ in units of $l_B \equiv GM/c_{s,out}$, $M$
the mass of the central black hole, and $c_{s,out}$ the isothermal
sound speed at the outer boundary.

\end{abstract}


\section{Introduction}

The amount of mass gravitationally accreted to the compact objects such
as black holes is determined by the conditions of gas around the
compact objects. The case when gas has a spherically symmetric
distribution, a polytropic pressure-density relation, and no angular
momentum was solved by Bondi (1952), who found that the mass accretion
rate for the transonic accretion is determined only by the density and
the temperature of surrounding gas, both assumed to approach constant
values far from the accreting objects. This rate, known as the Bondi
accretion rate, is widely used as `the mass accretion rate' in a
variety of accretion problems. As a function of the density
$\rho_\infty$ and the isothermal sound speed $c_{s,\infty}$ of gas at
infinity, the Bondi rate for pure hydrogen gas is
\begin{equation}\label{eq:Bondi_rate}
  \dot{M}_B
  =  4 \pi \Lambda \frac{(GM)^2 \rho_\infty}{\gamma^{3/2}c_{s,\infty}^3},
\end{equation}
where $G$ is the gravitational constant, $M$ the mass of the central
object, and $\gamma$ the adiabatic index of accreting gas. The constant
$\Lambda(\gamma)$ is 0.25 for $\gamma=5/3$ and 1.12 for $\gamma=1$. If
we define the Bondi radius as $r_B \equiv GM/c_{s,\infty}^2$, then
$\dot{M}_B = \Lambda \gamma^{-3/2} 4\pi r_B^2 \rho_\infty
c_{s,\infty}$. Gravity starts to dominate over gas pressure inside the
Bondi radius and the Bondi accretion rate is roughly the mass flux of
gas infalling into a sphere of radius $r_B$ with a velocity equal to
the sound speed at infinity.

However, many astrophysical accretion flows are expected to have
certain amount of angular momentum, and the angular momentum will
surely affect the flow properties, including the mass accretion rate.
Surprisingly, the rate of mass accretion for these rotating, viscous
accretion flows for given external boundary conditions, has not been
studied yet. We might expect that if the rotational support is
negligible at the Bondi radius, the effect of angular momentum to mass
accretion rate would be small. But the accretion rate would diminish as
the dimensionless measure of rotation $\lambda \equiv l_{out}/l_B$
approaches unity, where $l_{out}$ is the specific angular momentum of
gas at the outer boundary and $l_B \equiv r_B c_{s,\infty} $ is the
representative angular momentum expected at the Bondi radius.

Although the accretion problem can be reduced to local one in the limit
of thin accretion disc where the radial velocity is negligible (Shakura
\& Sunyaev 1973), accretion flow is fundamentally global in the sense
that the flow structure is only globally determined. As pointed out by
Yuan (1999), the flow property can be qualitatively affected by the
outer boundary conditions. The dependence of the mass accretion rate on
boundary conditions can only be addressed by constructing global
solutions. There have been many studies on the global solutions of
rotating accretion flow (e.g., Narayan, Kato, \& Honma 1997; Chen,
Abramowicz, \& Lasota 1997; Nakamura et al. 1997; Lu, Gu, \& Yuan 1999)
and the effects of outer boundary conditions on accretion flow (Yuan
1999; Yuan et al. 2000), but most have focused on the flow structures
and emission properties for a given mass accretion rate, and the
determination of how that rate is fixed by external circumstances
remains to be addressed.

In this work, we focus on the mass accretion rate of the accretion
flow, especially its dependence on the density, temperature, and the
angular momentum of gas at the outer boundary. We do this by
constructing the simplest global, transonic solutions within the slim
disk formalism (Abramowicz et al. 1988). Slim disk formalism uses
vertical integration to take into account the finite thickness of the
accretion disk, and more importantly allows for the radial motion of
the accretion flow and the critical point unlike the thin disk
approximation (Shakura \& Sunyaev 1973).

\section{Equations and Methods}

\subsection{Equations}

In slim disk approximation, the continuity equation is
\begin{equation}\label{eq:continuity}
  \dot{M} = - 4 \pi r H \rho v_r ,
\end{equation}
where $v_r$ is the radial velocity ($v_r < 0$ for inflow) and $\rho$
the density of the flow, both averaged over the height of the disk. The
disk scale height is $H \equiv c_s / \Omega_K$, where the isothermal
sound speed $c_s^2 \equiv P/\rho$, $P$ is the total gas pressure, and
$\Omega_K$ the Keplerian angular velocity for Paczy\'{n}ski-Wiita
potential (Paczy\'{n}ski \& Wiita 1980). This treatment allows for
general relativistic effect in an approximate fashion.

We assume that the gas is composed of hydrogen only and is fully
ionized. Ions and electrons are assumed to have the same temperature
$T$, and hence in this simplified treatment of a pure hydrogen gas, the
gas pressure is $P = (n_p + n_e) k T = 2 n_p k T = 2 (\rho/m_p) k T$,
where $n_p$ and $n_e$ are the number densities of protons and
electrons, respectively, $m_p$ the proton mass, and $k$ the Boltzmann
constant

The radial momentum equation is
\begin{equation}\label{eq:radial}
 v_r \frac{d v_r}{dr} + (\Omega_K^2 - \Omega^2) r + \frac{1}{\rho}
 \frac{dP}{dr} = 0,
\end{equation}
where $\Omega(r)$ is the angular velocity at radius $r$. Since we adopt
Paczy\'{n}ski-Wiita potential, the Keplerian angular velocity is
$\Omega_K^2 (r) \equiv GM r^{-3} [1-(r_{Sch}/r)]^{-2}$ where $r_{Sch}
\equiv 2GM/c^2$ is the Schwarzschild radius and $c$ the speed of light.

The angular momentum equation is given by
\begin{equation}\label{eq:angular}
 \rho v_r (\Omega r^2 - l_0) = \eta \alpha r P,
\end{equation}
where the constant $\alpha$ is the viscosity parameter of Shakura and
Sunyaev (1973): the viscous stress tensor $\ \tau_{\varphi r} = -
\alpha P$ (Abramowicz et al. 1988). Depending on exactly how the
viscosity description is implemented, the parameter $\eta$ can be from
$\eta = -2$ (Abramowicz et al. 1988), $\eta = -1$ (Nakamura et al.
1997), $\eta = (r/\Omega_K)(d\Omega/dr)$ (Narayan, Kato, \& Honma 1997)
to $\eta = d \ln \Omega_K /d \ln r$ (Narayan, Kato, \& Honma 1997; Yuan
1999). We choose $\eta = -1$ because this choice simplifies the angular
momentum equation into an algebraic one and has an added convenience of
automatically satisfying the no-torque condition at the black hole
horizon (Abramowicz et al. 1988, Yuan et al. 2000). On the other hand,
this implementation overestimates the shear stress when there is little
or no shear, and can cause difficulties when the solution is extended
to a very large radius. The integration constant $l_0$ is the specific
angular momentum accreted by the black hole, to be determined as the
eigenvalue during the construction of solutions for given boundary
conditions.

Finally, the energy equation is
\begin{equation}\label{eq:energy}
 \rho v_r \left[ \frac{d \epsilon}{dr} + P \frac{d}{dr}
 \big(\frac{1}{\rho}\big) \right] = q^{+} - q^{-},
\end{equation}
where $\epsilon = (\gamma-1)^{-1} P/\rho $ is the internal energy of
the gas per unit mass, $q^{+}$ and $q^{-}$ the heating and cooling
functions per unit volume, respectively. The only heating process
considered in the current treatment is the viscous heating, for which
we use the description
\begin{equation}\label{eq:visheating}
  q_{vis}^{+} = - \zeta \alpha P r (d\Omega/dr).
\end{equation}
We fix $\zeta = 1$ as in Abramowicz et al. (1988), Nakamura et al.
(1997), and Yuan et al. (2000). Slightly different values of $\zeta$
have been used in other works: for example, $\zeta= -
(r/\Omega_K)(d\Omega/dr)$ by Narayan, Kato, \& Honma (1997) and
$\zeta= - (d \ln \Omega_K / d \ln r)$ by Yuan (1999).

Since we mainly deal with high temperature accretion disk with
significant radial velocity, we use the optically thin, relativistic
ion-electron bremsstrahlung as the only cooling process,
\begin{equation}\label{eq:brems}
  q^{-} = \alpha_f r_e^2 m_e c^3 n_p n_e (32/3)(2/\pi)^{1/2}
  \left(\frac{kT}{m_e c^2}\right)^{1/2}
  \left[ 1 + 1.78 \left(\frac{kT}{m_e c^2}\right)^{1.34}\right],
\end{equation}
where $\alpha_f$ is the fine structure constant, $r_e$ the classical
electron radius, and $m_e$ the electron mass (Svensson 1982 and
references therein).

Equations (\ref{eq:radial}) and (\ref{eq:energy}) can be rearranged
into the form
\begin{equation}\label{eq:critform}
  \frac{d v_r}{dr} = \frac{A}{D}; \quad
  \frac{d c_s}{dr} = \frac{B}{D},
\end{equation}
where
\begin{eqnarray}\label{eq:ABD}
  A & \equiv & -\left( \frac{\gamma+1}{\gamma-1}
  + 2 \zeta \eta \alpha^2 \frac{c_s^2}{v_r^2} \right) v_r
  \left[ (\Omega_K^2 - \Omega^2) r
    + \frac{r c_s^2}{\Omega_K}\frac{d}{dr}
      \Big(\frac{\Omega_K}{r}\Big) \right] \nonumber \\
  &-& v c_s^2 \frac{r}{\Omega_K}\frac{d}{dr}\Big(\frac{\Omega_K}{r}\Big)
  - \alpha c_s^2 \left( 2 \zeta \frac{l_0}{r^2}
                 +\zeta\eta\frac{\alpha c_s^2}{r v_r} \right)
  + \frac{q^-}{\rho} \\
  B & \equiv & \left(1- \zeta \eta \alpha^2 \frac{c_s^2}{v_r^2}\right)
      c_s \left[ (\Omega_K^2 - \Omega^2)r
    + \frac{r c_s^2}{\Omega_K}\frac{d}{dr}
      \Big(\frac{\Omega_K}{r}\Big) \right] \nonumber \\
  &+& (v_r^2 - c_s^2) \left[ \frac{r c_s}{\Omega_K}
    \frac{d}{dr}\Big(\frac{\Omega_K}{r}\Big)
    + \alpha\frac{c_s}{v_r} \Big(2\zeta\frac{l_0}{r^2}
      +\zeta\eta\frac{\alpha c_s^2}{r v_r}\Big)
  - \frac{q^-}{\rho v_r c_s} \right] \\
  D & \equiv & \left(\frac{\gamma+1}{\gamma-1} + 2\zeta\eta \alpha^2
  \frac{c_s^2}{v_r^2}\right) v_r^2 -
  \left(\frac{2\gamma}{\gamma-1} + \zeta\eta \alpha^2
  \frac{c_s^2}{v_r^2}\right) c_s^2.
\end{eqnarray}
The zero of the denominator, $D=0$, yields the well-known critical (or
sonic) point condition $(v_r/c_s)^2 = 2\gamma/(\gamma+1)$ (see e.g.,
Narayan, Kato, \& Honma 1997), modified by viscosity. The exact value
of Mach number $\mathcal{M}_{cr} \equiv |v_r/c_s|_{r_{cr}}$ at the
critical point $r_{cr}$ is given by the root of the equation
\begin{equation}\label{eq:sonic}
  (\gamma+1){\mathcal{M}_{cr}}^4 - [2\gamma+2\alpha^2(\gamma-1)] {\mathcal{M}_{cr}}^2 +
  \alpha^2(\gamma-1) = 0
\end{equation}
for our specific choice of viscosity description. For $\alpha \ll 1$,
${\mathcal{M}_{cr}}^2 \approx 2\gamma (\gamma+1)^{-1}[1 + \alpha^2
(\gamma-1)(8\gamma-1)(4\gamma^2)^{-1}]$.

\subsection{Boundary Conditions and Method of Calculation}

For spherical accretion, the widely used Bondi accretion rate is solely
determined by the density and the temperature of accreting gas at
infinity. To be more precise however, there is no unique solution for a
given density and temperature at infinity: the same boundary condition
does allow subsonic (type I), transonic (type II), and unphysical (type
III) solutions with different mass accretion rates (Bondi 1952). The
additional requirement that the flow be transonic, i.e., regular at the
critical point, becomes an additional constraint, which uniquely
determines the solution and its accretion rate (cf. Parker 1963).

In rotating accretion flow, we expect one more physical quantity to be
important: the angular momentum of the gas. In the slim disk formalism,
the integration constant $l_0$ is an eigenvalue, and only a specific
value of $l_0$ for given $M$, $\dot{M}$, $\alpha$, $\rho_{out}$, and
$T_{out}$ admits a regular transonic solution (Muchotrzeb \&
Paczy\'{n}ski 1982, Abramowicz et al. 1988).

Spherical Bondi accretion has a critical point, which prevents the
direct integration of the equations from the outer subsonic region to
the inner supersonic region. This difficulty has been handled in two
ways: Either one starts with an arbitrarily chosen critical radius,
integrates both inward and outward with the help of regularity
conditions, and adjusts the critical radius until the outer and inner
(if any) boundary conditions are satisfied (Bisnovatyi-Kogan \&
Blinnikov 1980; Quataert \& Narayan 2000), or one starts with known
outer boundary conditions such as density and temperature of gas plus
initial choice of radial velocity (or equally mass accretion rate),
integrates inward toward the critical point, checks if the velocity
suddenly changes sign or diverges, adjusts the velocity (or the mass
accretion rate) accordingly until the integrated radial velocity
behaves normally even very close to the critical radius, extrapolates
across the critical point when the adjustment is precise enough, and
then integrates from there inward (Park 1990). This latter iteration
procedure has some resemblance to the physical behaviors of the
accretion flows. If the gas at outer boundary starts to accrete with
too small velocity (or mass accretion rate) than that of the transonic
solution, the gas flow stays subsonic all the way, never passing the
critical point (type I). On the other hand, if the gas starts to
accrete with too large velocity, the velocity of the flow diverges
before the critical point and becomes unphysical (type III). Only when
the gas starts to accrete with the just right velocity, or equally mass
accretion rate, can the flow pass through the critical point and become
steady transonic flow (type II).

The dynamics of rotating accretion flows are similar except that we
must include the added complications due to angular momentum and
viscosity. The flow must have the right mass accretion rate and angular
momentum for given density and temperature of gas at the outer boundary
to pass through the critical point. To find such regular transonic
solution, the angular momentum eigenvalue $l_0$ is searched by
iteratively integrating the equations, either from the outer boundary
(e.g., Muchotrzeb \& Paczy\'{n}ski 1982, Abramowicz et al. 1988,
Nakamura et al. 1997) or from the critical point in and out
(Chakrabarti 1996; Narayan, Kato, \& Honma 1997). Similar procedures
are also required to find the subsonic accretion solutions or
supersonic solutions with standing shocks onto compact objects with
hard surfaces, such as white dwarfs or neutron stars, that satisfy
specific outer and inner boundary conditions (Popham \& Narayan 1991;
Narayan \& Medvedev 2003).

Since we are more interested in the mass accretion rate of the
transonic flow for given boundary conditions at large radius, we
integrate from the outer boundary inward with the procedure used for
spherical accretion as explained above (Park 1990). The basic
parameters, the viscosity parameter $\alpha$ and the adiabatic index
$\gamma$ are chosen first. In this work, we assume $\gamma=5/3$ and
$\alpha=0.01$ unless noted otherwise. We then fix the density
$\rho_{out}$ and the temperature $T_{out}$ of the gas at the outer
boundary $r_{out}$. We also choose an integration constant $l_0$. We
start to integrate inward by arbitrarily choosing an initial guess for
the mass accretion rate $\dot{M}$, which, along with other conditions,
determines the radial velocity $v_r$ from equation
(\ref{eq:continuity}) and then the angular velocity $\Omega$ at the
outer boundary from equation (\ref{eq:angular}) to enable the initial
integration. The initial guess for $\dot{M}$ produces either a subsonic
solution or an unphysical, diverging solution, and integration stops.
We then adjust $\dot{M}$ until the integration can proceed as close to
the critical point as possible with a regular velocity profile. The
bifurcation between subsonic and diverging solution is quite sharp, and
the iterations can determine $\dot{M}$ up to an arbitrary precision.
Once the integration reaches close enough to the critical point, the
flow solutions are extrapolated across the critical point to the
supersonic region by an appropriate rational function, and the
integration ensues from therein. The interpolation introduces minimal
error because the supersonic part of non-transonic flow converges to
the transonic flow at smaller radii (see e.g., Bondi 1952 and Das
2007).

Since fixing $l_0$ and adjusting $\dot{M}$ is equivalent to fixing
$\dot{M}$ and adjusting $l_0$ and the transonic solutions exist only
for a rather limited range of $l_0$, we fix $\dot{M}$ and adjust $l_0$
in most cases. Moreover, equation (\ref{eq:angular}) relates $l_0$ to
the angular momentum at the outer boundary, $l_{out}$, and, therefore,
adjusting $l_0$ is also equivalent to adjusting $l_{out}\equiv
\Omega(r_{out}) r_{out}^2$. Since the reference angular momentum
against which $l_{out}$ needs to be compared is $l_B \equiv r_B c_s
(r_B) = GM/c_{s,out}$ which in fact is the Keplerian angular momentum
at $r_{B}$ if $r_{B} \gg r_{Sch}$, we define a dimensionless angular
momentum measure $\lambda = l_{out}/l_B$. To summarize, we construct a
global transonic solution for given $\rho_{out}$, $T_{out}$, and
$\lambda$.

Since our integration starts from a finite radius rather than from
infinity, we redefine the Bondi rate as $\dot{M}_B \equiv
4\pi\Lambda\gamma^{-3/2}(GM)^2 \rho_{out} c_{s,out}^{-3}$. With our
choice of cooling function, the whole problem can be put into a
dimensionless form, and the result, such as the dimensionless mass
accretion rate $\dot{m} \equiv \dot{M}/\dot{M}$, is independent of the
black hole mass $M$ if the density is scaled in units of
\begin{equation}\label{eq:rho0}
  \rho_0 \equiv \frac{m_p}{\sigma_{T} r_{Sch}} \simeq 1.7\times 10^{-5}
  \left(\frac{M}{M_\odot}\right)^{-1}~{\rm g}\,\,{\rm cm}^{-3},
\end{equation}
where $\sigma_{T}$ is the Thomson scattering cross section.

\section{Result and Discussion}

\subsection{Flow properties}

Figure 1 shows typical solutions with different amount of angular
momentum: high angular momentum (solid lines), intermediate angular
momentum (dotted lines), and low angular momentum (dashed lines). By
high angular momentum, we mean $\lambda \sim 1$, by low angular
momentum, $\lambda \ll 1$, and by the intermediate angular momentum,
between the two limits. The density and the temperature of gas at the
outer boundary are the same for the three solutions: $T_{out} =
2.75\times 10^{9} \K$ and $\rho_{out} = 1.5\times 10^{-6} \rho_0$ at
$r_{out} = 10^3 r_{Sch}$. The dimensionless angular momenta of gas at
the boundary and the mass accretion rate are: $\lambda = 1.7$ and
$\dot{m} = 0.05$ for high angular momentum flow; $\lambda = 0.27$ and
$\dot{m} = 0.386$ for intermediate angular momentum flow; $\lambda =
0.14$ and $\dot{m} = 0.643$ for low angular momentum flow. Most of the
thermal energy is advected with the flow rather than radiated away, and
temperature stays close to the virial value.

The flow with high angular momentum becomes supersonic at small radius
$r_{cr} = 2.2 r_{Sch}$ (solid line in Fig. 1a) and the angular momentum
(solie line in Fig. 1c) is close to or a few times lower than the
Keplerian value (short-long dashed line in Fig. 1c). The radial
velocity is significant but still smaller than the near free-fall
velocity of spherical Bondi flow. This is a typical hot, radiatively
inefficient advection-dominated accretion flow (ADAF) that has been
extensively studied (e.g., Narayan \& Yi 1994, 1995; Abramowicz et al.
1995; Narayan, Kato, \& Honma 1997; Chen, Abramowicz, \& Lasota 1997).

In spherical accretion, the flow passes the critical point at a much
larger radius $r_{cr} \sim r_B$, and the flow becomes supersonic inside
$\sim r_B$. But in rotating viscous accretion flow, a large part of the
flow is subsonic well inside $r_B$ because the rotation of the flow
balances out gravity, and the value of $r_{cr}$ depends on the angular
momentum of the flow. A low angular momentum flow (dashed line in Fig.
1a) has a critical point at a much larger radius $r_{cr} = 17 r_{Sch}$
compared to the high angular momentum one. It has a larger radial
infall velocity at the outer boundary as well because of smaller
centrifugal force (Fig. 1b), and the mass accretion rate is higher than
that of the high angular momentum flow. These characteristics show that
this low angular momentum flow has more resemblance to the spherical
flow than to the disk flow as is expected. This type of flow solution
is first discovered by Yuan (1999) and its dynamical and thermal
properties have been comprehensibly studied by Lu, Gu, \& Yuan (1999)
and Yuan et al. (2000).

The flow with intermediate angular momentum (dotted lines) has a
critical point at $r_{cr} = 3.2 r_{Sch}$, and shows intermediate
characteristics between the high and low angular momentum flow. The
flow properties change from disklike to quasi-spherical as the angular
momentum of the flow decreases. This is similar to the case of inviscid
rotating accretion flow, which becomes disklike or quasi-spherical,
depending on the angular momentum (Abramowicz \& Zurek 1981). This
transition of accretion flow from disklike to quasi-spherical in terms
of critical radius position is described in detail by Yuan et al.
(2000).

\subsection{Mass accretion rate}

In spherical accretion, the outer boundary is naturally selected to be
outside the critical point, where the density and the temperature of
gas stay roughly constant. In rotating viscous flow, the outer boundary
is not naturally associated with the critical point which is at much
smaller radius. So we consider two choices for the outer boundary
radius: one at the Bondi radius, $r_B(T_{out})$, for given $T_{out}$
and the other at a fixed radius regardless of $T_{out}$. The former
choice is equivalent to setting the outer boundary at a virial
temperature if we define the virial temperature as $T_{vir}(r) \equiv
GMm_p / (2 k r)$.

While original Bondi solutions are expressed in terms of the density
and temperature of gas at infinity, our solutions start at a finite
radius. But since the density and temperature of gas vary little from
infinity to the Bondi radius in the Bondi solutions, we compare our
solutions for given density and temperature of gas at finite $r_{out}$
against the Bondi solutions with the same density and temperature at
infinity.

We first discuss the case where the outer boundary is set to be the
Bondi radius, $r_{out} = GM/c_{s,out}^2 \propto T_{out}^{-1}$ for given
$T_{out}$, so that the solutions can be meaningfully compared with the
Bondi solutions. The density at the outer boundary is either
$\rho_{out} = 1.5 \times 10^{-6} \rho_0$ or $1.5 \times 10^{-9}
\rho_0$. Since the whole set of equations can be rescaled in density
except the cooling function $q^-$, the mass accretion rate then is
simply proportional to $\rho_{out}$ when cooling is not important. In
such case, the flow profile depends only on $\dot m$ and not on the
specific value of $\rho_{out}$. However, as the mass accretion rate
$\dot M$ approaches 0.1 times the Eddington mass accretion rate,
$\dot{M}_{Edd} \equiv L_{Edd}/c^2$, cooling becomes important as in
spherical accretion (Park 1990). In such high $\dot M/\dot{M}_{Edd}$
case, the existence and the properties of the flow depend on the
specific value of $\rho_{out}/\rho_0$.

Figure 2 shows the mass accretion rate of constructed solutions as a
function of the angular momentum at the outer boundary, in $\dot{m}$
versus $\lambda$ plane. Symbols represent the solutions for different
$T_{out}$: circles for $T_{out}=1.1\times10^{10} \K$
($r_{out}=2.5\times 10^2 r_{Sch}$), triangles for
$T_{out}=5.5\times10^9 \K$ ($r_{out}=5.0\times 10^2 r_{Sch}$), squares
for $T_{out}=2.8\times10^9 \K$ ($r_{out}=1.0\times 10^3 r_{Sch}$),
pentagons for $T_{out}=1.1\times10^9 \K$ ($r_{out}=2.5\times 10^3
r_{Sch}$), and hexagons for $T_{out}=5.5\times10^8 \K$
($r_{out}=5.0\times 10^3 r_{Sch}$). All these solutions are calculated
for $\rho_{out} = 1.5 \times 10^{-6} \rho_0$. The solutions for
$T_{out}=1.1\times10^7 \K$ ($r_{out}=2.5\times 10^5 r_{Sch}$),
temperature suitable for the ISM in a galactic nucleus, have much
larger $r_{out}$, and the bifurcation between subsonic to unphysical
branch becomes even sharper. Since the critical mass accretion rate
above which the hot, optically thin accretion disk does not exist also
decreases as $r_{out}$ increases (Abramowicz et al. 1995) and the Bondi
rate increases as $T_{out}$ decreases, we choose $\rho_{out} = 1.5
\times 10^{-9} \rho_0$ so that $\dot{M} /\dot{M}_{Edd}$ is in a range
where solutions can be found. The mass accretion rates of these
solutions are shown as crosses in the center of Figure 2.

We find that regardless of $T_{out}$, the mass accretion rate $\dot{m}$
decreases as the angular momentum $\lambda$ increases. It approaches
the Bondi accretion rate as $\lambda$ decreases. This change of the
mass accretion rate on the angular momentum is expected because given
density and temperature at $r_{out}$, $\rho$ and $H$ are fixed in the
continuity equation (Eq. [\ref{eq:continuity}]), and the mass accretion
rate is determined solely by the radial velocity $v_r$. From equation
(\ref{eq:radial}), the radial velocity is determined by the difference
between the gravity and the centrifugal accleration which is
proportional to $l^2/r^3$ (at $r \gg r_{Sch}$). Larger angular momentum
causes smaller infall velocity, and the mass accretion rate decreases,
and vice versa.

The mass accretion rate for the lowest angular momentum flow is quite
close to the Bondi rate ($\dot{m} \sim 1$) while that for the highest
angular momentum flow is roughly 20 times smaller than the
corresponding Bondi rate ($\dot{m} \sim 0.05$). We also find that
$\dot{m}$ as a function of $\lambda$ is rather insensitive to
$T_{out}$, or equally $r_{out}$ for $\dot{m} \ga 0.1$: The slope of
$\dot m(\lambda)$ has a tendency to become slightly flatter (from
circles to crosses) as $T_{out}$ decreases but the difference is not
large.

Since we expect the flow profile, especially the radial velocity, to
depend on the specific value of $\alpha$, we also calculate the
solutions for different values of $\alpha$. The mass accretion rates of
$\alpha=0.003$ are shown as a series of stars in the left side of
Figure 2 and that of $\alpha=0.03$ are shown in the right side of of
Figure 2. Compared to the solutions for $\alpha=0.01$ in the center
denoted by crosses, lower $\alpha$ flows have smaller mass accretion
rate and higher $\alpha$ flows larger mass accretion rate. Larger
$\alpha$ means stronger viscosity for given density and temperature,
which causes larger radial velocity, and hence larger mass accretion
rate, and vice versa for smaller $\alpha$. The relation between $\dot
m$ and $\lambda$  for $0.1 \la \dot{m} \la 1$ can be approximated by
\begin{equation}\label{eq:mdot_mB}
  \dot{m}(\lambda) = 0.09  \left( \frac{\alpha}{0.01} \right)
                       \lambda^{-1}.
\end{equation}
The three straight lines (from left to right) in Figure 2 show this
fitting function for $\alpha=0.003$, 0.01, and 0.03, respectively.

The mass accretion rate of the self-similar ADAF that has density
$\rho_{out}$ at $r_{out}$ has the same dependency on $\rho_{out}$ and
$r_{out}$ as the Bondi rate $\dot{M}_{B}$ (Narayan \& Yi 1994). Hence,
the dimensionless mass accretion rate of self-similar ADAF is simply
$\dot{m}_{ADAF} = 3.3 \alpha$. Since ADAF generally refers to $\lambda
\sim 1$ flow and $\dot{m}(\lambda=1) \simeq 3 \dot{m}_{ADAF}$, we find
that our solutions have approximately three times higher mass accretion
rates than those of self-similar ADAFs. The difference is probably due
to the fact that the density and velocity of global ADAFs do not
exactly follow the self-similar form, especially near the outer
boundary (Narayan, Kato, \& Honma 1997; Chen, Abramowicz, \& Lasota
1997).

For given $T_{out}$, the lower limit on the mass accretion rate is
determined by the Keplerian angular momentum barrier. A smaller mass
accretion rate demands smaller radial infall velocity. From equation
(\ref{eq:angular}), smaller $|v_r|$ requires larger $\Omega$, and
eventually $\Omega$ becomes larger than the Keplerian value $\Omega_K$
at $r_{out}$ for too small mass accretion rate. Gas cannot accrete in
steady-state when the angular momentum is much larger than the
Keplerian one, and therefore no solution exists when $\lambda \ga 3$.
Although we could determine $\dot m$ below $\dot{m} \simeq 0.1$ from
shooting until $\lambda$ approaches $\sim 1.5 - 3$ or 3, in many cases
we failed to construct the full transonic solutions down to the inner
boundary. The difficulty is much more severe for small $\alpha$: the
bifurcation between subsonic and unphysical occurs at a large radius
and computing accuracy is not enough to integrate all the way down to
the horizon. So we have limited confidence in the solutions below
$\dot{m} \la 0.2$ for $\alpha = 0.003$ and 0.03. The upper limit on the
mass accretion rate on the other hand is determined by the Bondi
accretion rate itself. The lowest angular momentum flow has its mass
accretion rate already quite close to the Bondi rate, which is the mass
accretion rate for zero angular momentum flow. Thus any higher mass
accretion flow will end up on the unphysical branch of solutions in
spherical Bondi flow (type III), with minimum effect due to the angular
momentum.

We can think of extending the velocity versus radius diagram of Bondi
(1952) to a similar one with the angular momentum effect added. What
angular momentum does is to lower the critical mass accretion rate for
the transonic flow. As the angular momentum of the flow increases, the
critical mass accretion rate decreases, and the rotating accretion flow
accretes transonically at a critical mass accretion rate lower than the
original Bondi accretion rate. On the other hand, if the angular
momentum of the flow decreases, the critical mass accretion rate
increases up to the Bondi accretion rate which is the critical mass
accretion rate for zero angular momentum flow, above which no transonic
steady-state rotating accretion flow exists. Since equation
(\ref{eq:mdot_mB}) gives $\dot{m} > 1$ for $\lambda < 9 \alpha$ and
$\dot{m}$ cannot be greater than unity, it looks like there is no
solution for $\lambda < 9 \alpha$. However, we are using slim disk
formulation with a simple viscosity prescription to describe what is
really a two-dimensional quasi-spherical flow, especially for lowest
angular momentum flow; so the non-existence of solutions below the
lower limit on $\lambda$ is probably caused by the approximate physical
descriptions of the flow, especially the overestimated viscosity for
the low or no angular momentum flow. Surely, we will have a spherical
Bondi flow when the angular momentum is zero. Therefore, we expect that
steady flow probably exists even below $\lambda \sim 9 \alpha$ and
\begin{equation}\label{eq:mdot_zerolambda}
  \dot{m} \simeq 1
  \quad {\rm for} \quad \lambda \la 9 \alpha .
\end{equation}
We conclude that the steady-state, transonic hot accretion flow exists
only when the mass accretion rate is within a certain range or the
angular momentum at the outer boundary is below a certain maximum
value. For our specific choice of parameters and conditions, the range
in the mass accretion rate is
\begin{equation}\label{eq:mdot_range}
  \dot{m}_{cr} \la \dot{m} \leq  1,
\end{equation}
where $\dot{m}_{cr}=\dot{m}(\lambda_{cr})$ is the mass accretion rate
of the flow with maximum possible angular momentum at the outer
boundary. Our calculations suggest $\lambda_{cr} \la 3$ and
$\dot{m}_{cr} \sim 0.05$ for $\alpha = 0.01$.

Now we discuss flows with the outer boundary temperature different from
the virial temperature. For example, when the cool, outer geometrically
thin disk is attached to the hot inner accretion disk, the temperature
at the outer boundary will be much lower than the virial temperature
(Narayan, Kato, \& Honma 1997). So we fix $r_{out} = 10^3 ~r_{Sch}$ but
vary $T_{out}$. The outer boundary density is the same at $\rho_{out} =
1.5 \times 10^{-6} \rho_0$. Figure 3 shows the mass accretion rate
again in $\dot{m}$ versus $\lambda$ plane. Symbols are the same as in
Figure 2, except circles for $T_{out}=3.6\times10^9 \K$, triangles for
$T_{out}=2.2\times10^9 \K$, squares for $T_{out}=1.1\times10^9 \K$,
pentagons for $T_{out}=5.5\times10^8 \K$, and hexagons for
$T_{out}=2.0\times10^8 \K$.

Again, solutions exist only in a limited range of $\dot{m}$, or equally
$\lambda$. But this time both upper and lower limits on $\dot{m}$ vary
with $T_{out}$. The reason for the difference is that $r_{out}$ is
fixed and not related to $T_{out}$ whereas it is related to $T_{out}$
in Bondi solutions. As in previous choice of boundary, the lower limit
on $\dot{m}$ is from the Keplerian angular momentum barrier. However,
the upper limit is determined for a different reason in this case. As
the mass accretion rate for given boundary conditions increases, the
radial infall velocity at the outer boundary should increase. But the
increasing radial infall velocity eventually becomes larger than the
sound speed at the boundary. The whole flow becomes supersonic from the
outer boundary to the innermost radius. Since we are looking for
transonic solutions that uniquely determine the mass accretion rate, we
disregard these supersonic branch of solutions, and no transonic
solutions exist above certain upper limit on $\dot{m}$. This upper
limit on $\dot{m}$ decreases as $T_{out}$ decreases because the sound
speed decreases with $T_{out}$ and the flow becomes supersonic at
smaller radial velocity, or mass accretion rate, causing rapid decrease
of the upper limit on $\dot{m}$ with decreasing $T_{out}$.

The dependence of $\dot{m}$ on $\lambda$ is also different. Again,
$\dot{m}$ decreases as $\lambda$ increases, but is not inversely
proportional to  $\lambda$, nor is independent of $T_{out}$. Cooler
flow, i.e., lower $T_{out}$, has lower mass accretion rate $\dot{m}$
for a given $\lambda$. When $\lambda = 0.1$, $\dot{m}$ for
$T_{out}=2.2\times10^9 \K$ is $\sim 1$ while that for
$T_{out}=2.0\times10^8 \K$ is as low as $\sim 10^{-4}$. This means that
if gas is injected into a given radius with near-Keplerian angular
momentum but with temperature much below the virial temperature at that
radius, the mass accretion rate can be a few orders of magnitude below
the Bondi accretion rate.

\section{Summary}

We have constructed global solutions of the rotating, viscous accretion
flow within the framework of height-averaged, slim $\alpha$ disk with
varying amount of specific angular momentum $\lambda$ at the outer
boundary. We find that the low angular momentum flow, the flow with
$\lambda \ll 1$, resembles the spherical Bondi flow and its mass
accretion rate approaches the Bondi accretion rate for the same density
and temperature at the outer boundary. The high angular momentum flow,
the flow with $\lambda \sim 1$, on the other hand is the conventional
hot accretion disk with advection, but its mass accretion rate can be
significantly smaller than the Bondi accretion rate for the same
density and temperature at the outer boundary.

We also find that solutions exist only within a limited range of the
dimensionless mass accretion rate $\dot{m}$, the mass accretion rate in
units of the Bondi accretion rate. When the temperature at the outer
boundary is equal to the virial temperature at that radius, solutions
exist only when $ \dot{m}_{cr} \la \dot{m} \leq 1$ because below
$\dot{m}_{cr}$ the angular momentum of the flow becomes higher than the
Keplerian value, and the steady accretion is not possible. The exact
value of $\dot{m}_{cr}$ as a function of the viscosity parameter
$\alpha$ is not known, but is $\sim 0.05$ for $\alpha = 0.01$. The
upper limit on the mass accretion rate is the Bondi accretion rate
since it is the mass accretion rate of zero angular momentum flow.
Moreover, we also find that the dimensionless mass accretion rate is
roughly independent of the radius of the outer boundary but inversely
proportional to the angular momentum at the outer boundary in units of
the Keplerian angular momentum at that same radius and proportional to
the viscosity parameter, $\dot{m} \simeq 9.0\ \alpha \lambda^{-1}$ when
$0.1 \la \dot{m} \la 1$. When the temperature at the outer boundary is
much less than the virial temperature, the mass accretion rate can be a
few orders of magnitude smaller than the Bondi accretion rate.

The fact that the mass accretion rate can be smaller than the Bondi
accretion rate and depends on the angular momentum of gas at the outer
boundary could have interesting implications for many astrophysical
accretion systems, especially underluminous AGNs and galactic centers.

Although we now have a better understanding of the behavior of
rotating, viscous accretion flows with varying amount of angular
momentum, the precise numerical values presented in this work are
expected to contain some uncertainties because these hot accretion
flows are certain to have real two-dimensional structures, and the
one-dimensional, height-averaged approximation adopted in this work
will only approximately describe the real flow. Hence, more precise
analyses of these accretion flow including other relevant physics such
as shocks (Chakrabarti \& Das 2004), outflows (Stone, Pringle \&
Begelman 1999), or radiative heating (Park \& Ostriker 1999, 2001,
2007; Yuan, Xie \& Ostriker 2009; Proga, Ostriker, \& Kurosawa 2008)
will require careful multi-dimensional studies (see e.g. Proga 2007,
Kurosawa \& Proga 2008, and Proga, Ostriker, \& Kurosawa 2008 for
current multi-dimensional studies).

\acknowledgments

We thank Jerry Ostriker for many insightful discussions and careful
reading of the manuscript. We also thank the referee, Ramesh Narayan,
for useful comments which significantly improved this paper. This work
was supported by the Korea Research Foundation Grant funded by the
Korean Government(MOEHRD) (KRF-2007-013C00028) and also by the National
Research Foundation of Korea(NRF) grant funded by the Korea
government(MEST) (No. 2009-0062868).

\clearpage

\begin{figure}
\plotone{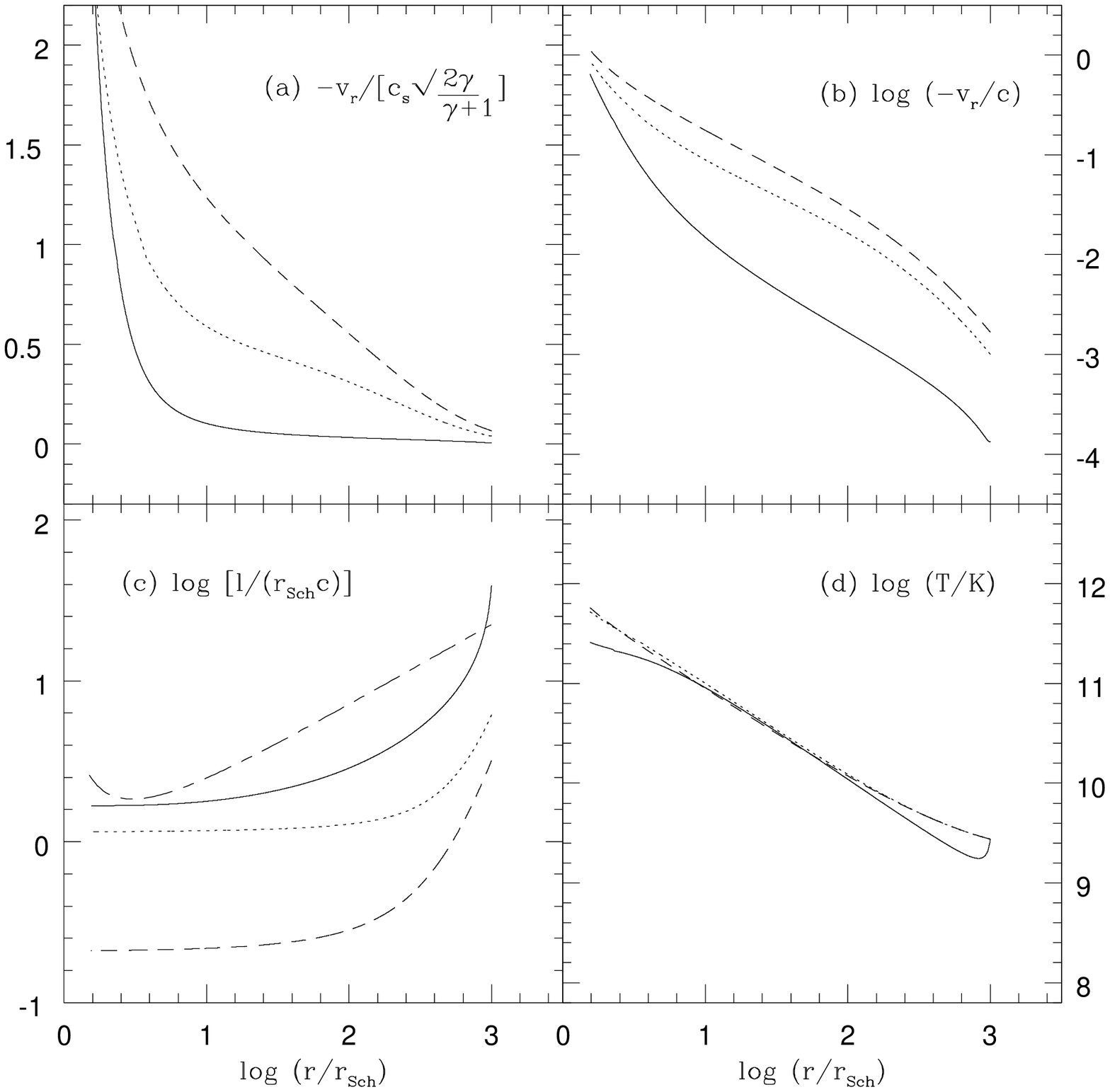} \caption{The (a) Mach number, (b) velocity, (c)
angular momentum, and (d) temperature profiles of typical high (solid
lines), intermediate (dotted lines), and low angular momentum (dashed
lines) solutions. Long-short dashed line in (c) is the Keplerian
angular momentum for Paczy\'{n}ski-Wiita potential. See text for
details.}
\end{figure}

\begin{figure}
\plotone{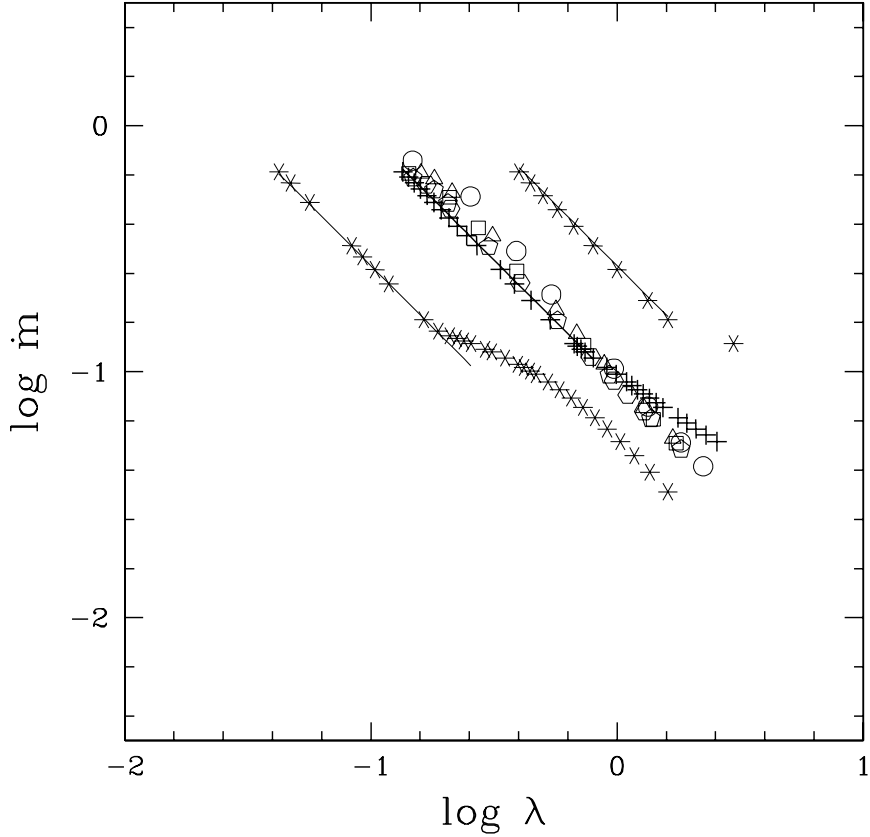} \caption{The mass accretion rate in units of the Bondi
rate as a function of the angular momentum at the outer boundary in
units of $l_B$ for an outer boundary at the Bondi radius when the
temperature at the outer boundary is equal to the virial temperature.
Different symbols represent different temperature at the outer
boundary: circles for $T_{out}=1.1\times10^{10} \K$, triangles for
$T_{out}=5.5\times10^9 \K$, squares for $T_{out}=2.8\times10^9 \K$,
pentagons for $T_{out}=1.2\times10^9 \K$, hexagons for
$T_{out}=5.5\times10^8 \K$, and crosses for $T_{out}=1.1\times10^{7}
\K$, but all for $\alpha=0.01$. A series of star symbols on the left
represent $T_{out}=1.1\times10^{7}\K$ and $\alpha=0.003$ and those on
the right $T_{out}=1.1\times10^{7}\K$ and $\alpha=0.03$, respectively.}
\end{figure}

\begin{figure}
\plotone{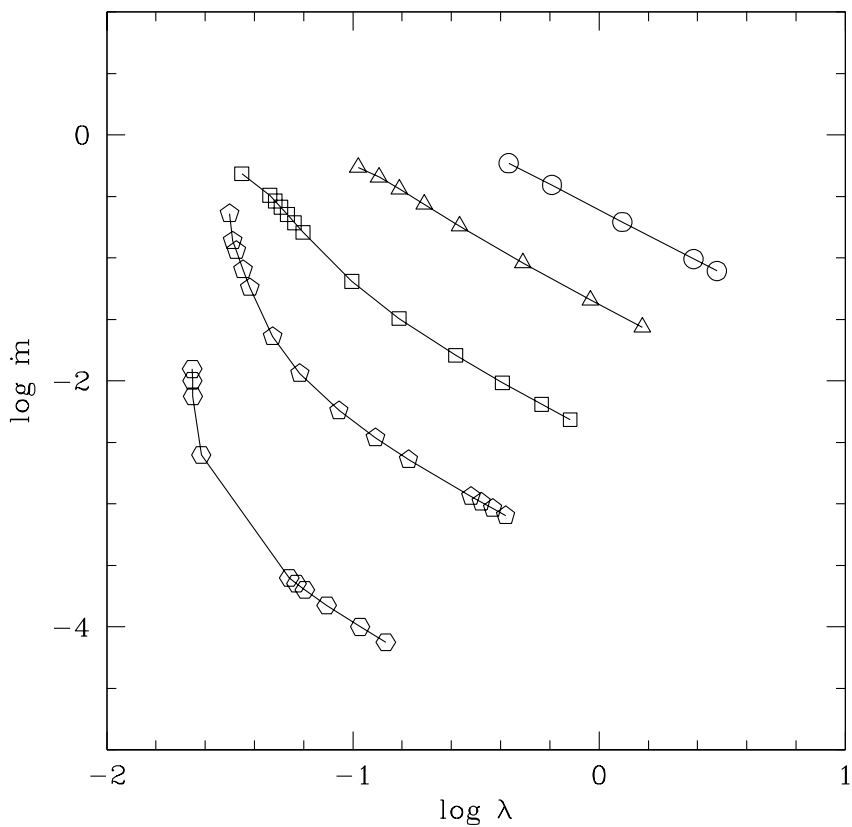} \caption{The mass accretion rate in units of the Bondi
rate as a function of the angular momentum at the outer boundary in
units of $l_B$ for a fixed outer boundary at $r_{out}= 10^3 r_{Sch}$
and $\alpha=0.01$. Symbols represent the same as in Figure 2, except
circles for $T_{out}=3.6\times10^9 \K$, triangles for
$T_{out}=2.2\times10^9 \K$, squares for $T_{out}=1.1\times10^9 \K$,
pentagons for $T_{out}=5.5\times10^8 \K$, and hexagons for
$T_{out}=2.0\times10^8 \K$.}
\end{figure}

\end{document}